\documentclass[10pt,journal,compsoc]{IEEEtran}
\IEEEoverridecommandlockouts
\usepackage[linesnumbered,ruled,vlined]{algorithm2e}
\usepackage[justification=centering]{caption}
\usepackage{dblfloatfix}   
\usepackage{url}

\usepackage{breakurl}
\usepackage[breaklinks]{hyperref}
\usepackage{diagbox,threeparttable,mwe} 
\usepackage[numbers,sort&compress]{natbib}
\usepackage{amsmath,amssymb,enumitem,esvect}
\usepackage{algorithmic}
\usepackage{graphicx, wrapfig}
\usepackage{textcomp}
\usepackage{xcolor, comment, setspace,empheq}
\newtheorem{defn}{Definition}
\newcommand{\M}{\mathcal{M}}
\newcommand{\E}{\mathbb{E}}
\newcommand{\V}{\mathbb{V}}

\allowdisplaybreaks
\def\BibTeX{{\rm B\kern-.05em{\sc i\kern-.025em b}\kern-.08em
    T\kern-.1667em\lower.7ex\hbox{E}\kern-.125emX}}

\begin{document}

\title{Privacy-preserving Travel Time Prediction with Uncertainty Using GPS Trace Data\\
\author{Fang Liu$^*$,  Dong Wang$^*$,  Zhengquan Xu
\thanks{$^*$Co-first authors.  F. Liu is Professor in the Department of Applied and Computational Mathematics and Statistics, University of Notre Dame, Notre Dame, IN, 46556, USA  (corresponding author; e-mail: Fang.Liu.131@nd.edu); D. Wang  (e-mail: dwang22@nd.edu)  is a doctoral student and Z. Xu  (e-mail: xuzq@whu.edu.cn) is Professor in the Department of State Key Laboratory of Information Engineering
in Surveying, Mapping and Remote Sensing, Wuhan University, Wuhan, 430079, China.}\vspace{-18pt}
}}
\maketitle
\vspace{-36pt}
\begin{abstract}
The rapid growth of GPS technology and mobile devices has led to a massive accumulation of location data, bringing considerable benefits to individuals and society. One of the major usages of such data is travel time prediction, a typical service provided by GPS navigation devices and apps.  Meanwhile, the constant collection and analysis of the individual location data also pose unprecedented privacy threats. We leverage the notion of geo-indistinguishability, an extension of differential privacy to the location privacy setting, and propose a procedure for privacy-preserving travel time prediction without collecting actual individual GPS trace data. We propose new concepts to examine the impact of geo-indistinguishability-based sanitization on the usefulness of GPS traces and provide analytical and experimental utility analysis for privacy-preserving travel time prediction. We also propose new metrics  to measure the adversary error in learning individual GPS traces from the collected sanitized data. Our experiment results suggest that the proposed procedure provides travel time prediction with  satisfactory accuracy at reasonably small privacy costs.
\end{abstract} \vspace{-3pt}

\begin{figure*}[!t]
\centering \includegraphics[width=1\textwidth]{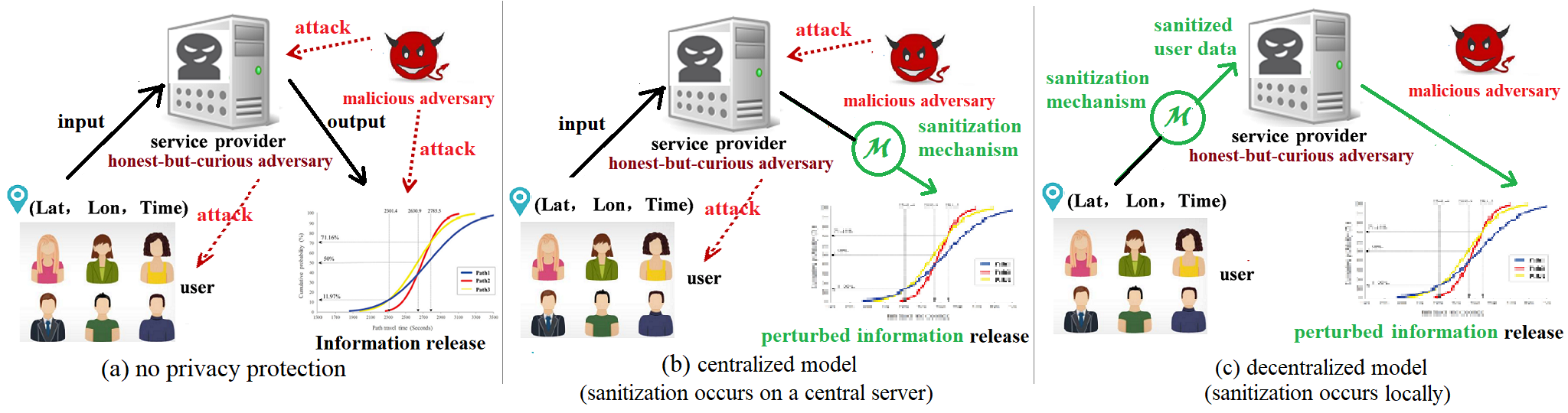}
\vspace{-18pt}
\caption{Privacy Protection Strategies for Analysis using GPS Data }\label{fig:adversary}\vspace{-12pt}\end{figure*}  

\vspace{-12pt}
\section{Introduction}\label{sec:intro}\vspace{-3pt}
\subsection{Motivation and Problem}\label{sec:problem}
\vspace{-3pt}
The rapid growth of GPS technology and mobile devices has led to a quick accumulation of massive location data. Analysis and understanding of the data have brought enormous benefits to individuals and society.  Meanwhile, collection and processing of location data can easily expose personal behaviors, interests, social relations, or other private information, especially if combined with other data sources. \citet{de2013unique} studied 15-month location data from 1.5 million people and found that as little as 4 space-time points can uniquely identify 95\% individuals.  Meanwhile, users are often not fully aware of privacy risks from sharing their location data with service providers and how their data are used \cite{coppens2014privacy, kessler2018geoprivacy}.

One  important application of GPS data is Travel time Prediction with Uncertainty (TPU). TPU examines how quickly a person arrives at a destination with a certain level of confidence. It is important for transportation and urban planning  and a typical route planning service provided by navigation systems and mapping apps.  TPU often relies on continuous collection and processing of users' travel trajectories and  thus exposes data contributors to privacy threats from adversaries, honest-but-curious (e.g., the service provider itself) and malicious, as depicted in Fig. \ref{fig:adversary}(a)). 

To our best knowledge,  there is no work focusing on privacy-preserving TPU (PP-TPU) analysis. We display in Fig.~\ref{fig:adversary}(b) and \ref{fig:adversary}(c) two possible strategies for PP-TPU. Fig. \ref{fig:adversary}(b) focuses on sanitizing aggregated statistics calculated from the actual user data, say via a differentially private randomization mechanism such as the Laplace \cite{dwork2006calibrating} or the Gaussian mechanisms \cite{dwork2014algorithmic,liu2018generalized}). This strategy  mitigates the privacy threats from the adversaries who aim to learn something new  about their targets from the released aggregate information, but it cannot manage the privacy risk brought by the adversaries who have access to the original data, such as the service provider itself. In Fig.~\ref{fig:adversary}(c), sanitization occurs during data collection; that is, the true individual responses go through a sanitization mechanism locally before being shared with a third party. As a result, the true responses are only known to the users themselves. 

We aim to develop a PP-TPU procedure that  implements the strategy in Fig.~\ref{fig:adversary}(c), leveraging the state-of-the-art notions and sanitization mechanisms in data privacy research as stepping stones to achieve our goal. 

\vspace{-12pt}\subsection{Related Work}\vspace{-6pt}
Data encryption, anonymization, and obfuscation are common frameworks for controlling the privacy risks incurred by location data collection and sharing. The PP-TPU procedure we propose can be regarded as a data obfuscation approach. Below we provide a brief overview of each framework,  analyze their limitations and challenges, and state the rationale for us adopting the data obfuscation framework to develop the PP-TPU procedure. 

Location encryption uses cryptographic techniques to mitigate privacy risks in location data \cite{zhong2007louis, popa2011privacy, zhu2011applaus, mascetti2011privacy, li2019privacy,zhu2019traffic, zhang2019privacy}. This is a common approach for protecting individual privacy but can be costly in terms of computation and resource \cite{zhou2019privacy}. Furthermore,  data, once decrypted,  are no longer private to those who have the authority to access the data; the privacy:utility ratio from the data user perspective is either 100:0\% or 0:100\%, corresponding to the two states of encryption and decryption, respectively.  These two extreme options of data access often do not meet the practical needs for data sharing. Indeed, a non-zero small privacy cost is often acceptable in practice so to create more options between the two extremes and  share information with more data users.

Data anonymization and obfuscation  provide options between the two extremes. These concepts focus on privacy-preserving data processing and analysis via methods such as data coarsening, removal of identifiers,  reporting dummy locations via randomization mechanisms, among others.  The key issue in these approaches is to strike a good balance between privacy loss and data utility (the higher the privacy loss, the more utility there is in the anonymized or obfuscated data relative to the original data).  

Several formal privacy concepts have been developed to attain  anonymization for general data, such as $k$-anonymity  \cite{samarati2001protecting, sweeney2002k} and $l$-diversity \cite{machanavajjhala2007diversity}. Both concepts have been adapted and applied in the location privacy setting (e.g.,  \cite{gruteser2003anonymous,kido2005protection,gedik2007protecting, ghinita2007prive,lu2008pad,chow2009casper,gkoulalas2010providing,niu2014achieving,fei2017k,zhao2018illia,wang2019achieving, zhang2019caching} for $k$-anonymity, \cite{terrovitis2008privacy, liu2009query, tu2018protecting} for $l$-diversity). Though the concepts are intuitive, neither $k$-anonymity nor $l$-diversity  involves randomization; it has been shown that adversaries may still learn sensitive information or re-identify individuals from anonymized location data \cite{de2013unique,wang2018privacy}. 

Differential privacy (DP) \cite{dwork2006calibrating} involves randomization and is a mathematically robust conceptual for data obfuscation. DP bas been proved to be robust against a wide range of adversary attacks \cite{nissim2017differential, dwork2017exposed}, processes properties such as privacy loss composability \cite{mcsherry2007mechanism} and immunity to post-processing \cite{dwork2014algorithmic} that facilitate practical implementation, and has quickly become the mainstream in privacy research and applications. Big tech companies (e.g., Apple, Google, IBM) and government agencies (Census2020) also adopt DP or its variants to collect or release data. The classical DP concept is also applicable in location data analysis \cite{asada2019and, bavadekar2020google, aktay2020google, xiao2015protecting}.

Conceptual extensions of DP for location privacy also exist, among which  geo-indistinguishability  (GI) \cite{andres2013geo} is perhaps the most popular. GI has been explored in a wide range of location privacy applications.  \citet{andres2013geo} propose the planar Laplace mechanism  to perturb the 2-dimensional location data.    \citet{chatzikokolakis2015constructing} extend GI by developing an elastic indistinguishability metric that adapts the amount of injected noises according to the area density. \citet{cunha2019clustering} propose a clustering  mechanism for continuous location traces clustering application.  \citet{shi2019deep} apply  GI to preserve location information of passengers in vehicles of transportation companies.   \citet{qian2020privacy} propose a GI task allocation  mechanism to preserve location privacy in mobile crowdsensing applications. \citet{qiu2020location} design a strategy to minimize the loss in quality of service  due to GI obfuscation. \citet{shi2020quantitative} present a closed-form relationship between localization accuracy and the  GI privacy level. \citet{takagi2020poster} propose  Geo-Graph-Indistinguishability that extends DP to the setting of location privacy on road networks. \citet{ren2020egeoindis} present the ``Expanding GI'' framework to protect the privacy of vehicle locations by abstracting maps as bitmaps and utilizing linear programming to control information loss. 

In summary, despite the various applications of GI in  practical problems, its potential in TPU analysis has not been explored to our knowledge. The reasons for us adopting the GI concept for developing a PP-TPU procedure rather than other  privacy protection frameworks for location data are as follows.  First, our GI-based PP-TPU procedure avoids the limitations of encryption-based approaches,  thanks to the tunable parameter (privacy budget/loss $\epsilon$) in GI, and provides options between the two extremes of zero utility/full protection ($\epsilon\!\rightarrow\!0$) and full utility/zero protection ($\epsilon\!\rightarrow\!\infty$). Second, while PP-TPU can leverage the classical DP concept to release aggregate location information, it often requires the collection of actual individual data (Fig.~\ref{fig:adversary}(b)). In contrast, GI provides a framework to perform TPU analysis without collecting  actual individual GPS records (Fig.~\ref{fig:adversary}(c)). This strategy, compared to Fig.~\ref{fig:adversary}(b), better protects  users' privacy and boost users' confidence in data sharing  as the data sanitization is performed locally on users' own devices; in addition, users can pick $\epsilon$ themselves depending on how willing they are to share their data. Third, GI follows similar mathematical reasoning as DP and many desirable properties of DP are also applicable to GI such as composability and immunity to post-processing that are important to meeting the specific challenges in the development of PP-TPU procedures.  Specifically, the TPU analysis requires the collection of multiple 3-tuple GPS records (2-dimensional location coordinates and timestamp)  from a single trip in a traveler. Due to the sequential composition of privacy loss, the overall privacy loss can become unrealistically high  to obtain a useful sanitized trajectory, or the sanitized trajectory is useless if the overall cost is kept reasonable. Besides, the GPS records, after sanitization, have to make sense in the 3-dimensional spatiotemporal space and for the actual road networks on which TPU is performed. 

\vspace{-9pt}\subsection{Our Contribution}\vspace{-3pt}
Our PP-TPU procedure balances the trade-off between the privacy risk from collecting  individual travel trajectories and the utility of sanitized trajectories, while being mindful of its practical feasibility. The conceptual and methodological contributions and the potential practical impacts of the procedure are summarized as follows.
\vspace{-3pt}
\begin{itemize}[leftmargin=9pt]
\item The procedure takes into account privacy loss composability when sanitizing multiple GPS records per trajectory. To limit the overall privacy loss, it curbs the number of GPS records collected per trajectory and leverages public information of maps to filter out unusable trajectories.
\item  We propose new concepts of \emph{usable set of travel trajectories}, \emph{effective number of mapped full trajectories}, \emph{usefulness}, and different distance deviation measures to quantify the utility of sanitized GPS records and trajectories for PP-TPU.  
\item  We propose new metrics \emph{continuous positioning degree} and \emph{average distance} to quantify the adversary error in learning individual trajectories given sanitized GPS records.
\item We examine the feasibility of the proposed PP-TPU procedure by quantifying the trade-off between privacy loss and the utility of sanitized GPS records and traces analytically and empirically, providing insights on choosing privacy loss parameters in different application scenarios.
\item The  procedure is easy to implement. Service providers and GPS navigation systems and apps may use it to collect user location data and provide TPU service with guaranteed privacy protection.
\end{itemize}

\vspace{-9pt}\section{Preliminaries}\label{sec:prelim}
\vspace{-3pt} \subsection{Map Matching of GPS Records}
\begin{figure}[!htb]\vspace{-9pt}
\centering \includegraphics[width=0.45\textwidth]{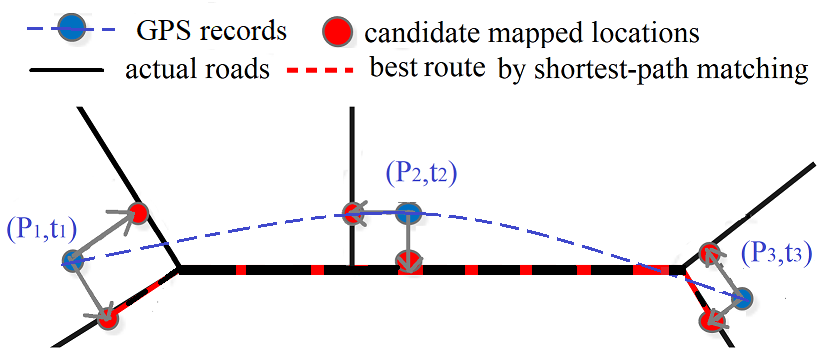}\vspace{-3pt}
\caption{Shortest path map matching}\label{fig:gps mapping}
\end{figure}

\vspace{-3pt}
TPU starts with the collection of GPS records that contain the spatial-temporal information of a traveler and then matches the GPS locations with physical road network maps. Each GPS record contains the location  $P_i$ (latitude and longitude coordinates) and timestamp $t_i$ information. Due to satellite signal blockage, multi-path effects, and other factors that may affect GPS signals,  collected GPS location information is not always accurate.  Commonly direct projections of GPS coordinates may not correspond to any meaningful real map coordinate, and road mapping algorithms often involve some level of approximation.  Fig. \ref{fig:gps mapping} shows an example of the shortest path map-matching algorithm  \cite{orda1990shortest} that project 3 GPS registration points onto the physical road network.
\vspace{-15pt}   

\subsection{Travel Time Prediction with Uncertainty}\vspace{-3pt}
Various approaches to travel time prediction have  been  developed. The na\"ive travel time prediction outputs a single projected travel time value, but the reachable time-space range of a traveler is rather stochastic, due to the dynamic nature of  human  behaviors, traffics, etc \cite{noland1995travel}. Studies \cite{chen2017most} have shown that individuals, when facing uncertainty in travel time,  tend to avoid the risk of lateness and often reserve some time to ensure that they can arrive on time with a high level of confidence \cite{de2013unique}. It is also important for transportation and urban planning to understand the uncertainty around travel time for infrastructure development and designs, among others. 

The analysis of TPU aims to obtain $f(t)$, the probability density function (pdf) of travel time $t$ spent over a trip with starting point A and destination B.  The probability that the destination B can be reached within time $b$ can be easily obtained from the cumulative density function (CDF) of $t$, that is, $p\!=\!F_t(b)\!=\!\int_{0}^{b}f(t)dt$. For example, suppose $b\!=\!10$ minutes and $p\!=\!0.9$, then there is a 90$\%$ chance of arriving at the destination within 10 minutes. In reality, $f(t)$ is unknown and is often estimated by an empirical $\hat{f}(t)$ based on collected individual travel data, such as via the GPS.

\vspace{-6pt}\subsection{Differential Privacy and Geo-indistinguishability}\vspace{-3pt}
Differential privacy (DP) is a state-of-the-art privacy protection model that guarantees privacy for released information in mathematically rigorous terms. 
\begin{defn}[\textbf{$\epsilon$-differential privacy} \cite{dwork2006calibrating}]\label{def:DP}
A randomization mechanism $\M$ is $\epsilon$-differentially private, if for any pair data sets $X$ and $X'$ that differ by one record and every possible outcome set $\Omega$ to a query,
\begin{equation}
\Pr[\M(X)\in\Omega]\le{e^{\epsilon}}\cdot{\Pr[\M(X')\in{\Omega}]},
\end{equation}
\end{defn}\vspace{-6pt}
where  $\epsilon>0$ is the privacy budget or loss parameter. The smaller $\epsilon$ is, the more privacy protection there is on the individuals in the data set. $X$ and $X'$ differ by one record may refer to the case that $X$ and $X'$ are of the same size but differ in the attribute values in exactly one record, or the case that $X'$ is one record more than $X$ or vice versa. 

The classical DP in Definition \ref{def:DP} provides a mathematical model for privacy guarantees when releasing  aggregate statistics from a group of individuals. The local DP \cite{localDP2011, duchi2013local} is an extension of the classical DP to a single user's data and can be used to develop mechanisms for sanitizing individual responses rather than aggregate results, with privacy.
\begin{defn}[\textbf{$\epsilon$-local differential privacy} \cite{kasiviswanathan2011can, duchi2013local}]\label{def:localDP} 
A randomization mechanism $\M$ provides $\epsilon$ local DP if
\begin{equation} 
\Pr[\M(x)\in\Omega]\le{e^{\epsilon}}\cdot{\Pr[\M(x')\in{\Omega}]}.
\end{equation}
for all pairs of possible data $x$ and $x'$ from an individual and all possible output subset $\Omega$ from $\M$.
\end{defn}
The local DP implies that even if an adversary has access to the  sanitized  personal responses from a randomization mechanism that satisfies local DP, the adversary is still unable to learn much new about the user's actual responses. 

GI is an extension of DP to location privacy and aims at releasing individual location records. In that sense, GI is more similar to the local DP concept than the classical DP; but all three concepts are based on similar mathematical formulations.  The formal definition of GI is given below.
\begin{defn}[\textbf{Geo-indistinguishability} \cite{andres2013geo}]\label{def:GI}
Let $d(P,P')$  denote the Euclidean distance between any two distinct locations $P$ and $P'$, and $\epsilon$ be the unit-distance privacy loss. A randomization mechanism $\M$ satisfies GI iff for all possible released location $P^*$, $\gamma>0$, and any possible pair of $P$ and $P'$ within the radius of $\gamma$, 
\begin{equation}\label{Eq:geoDP}
\Pr(\M(P)=P^*|P)\le e^{\epsilon\gamma}\cdot\Pr(\M(P')=P^*|P').
\end{equation}
\end{defn}\vspace{-3pt}
In other words, $\M$ in Eq (\ref{Eq:geoDP}) enjoys $(\epsilon\gamma)$-privacy for any specified $\gamma$, and  the probability of distinguishing any two locations with a radius of $\gamma$, given the released location $P^*$, is $e^{\epsilon\gamma}$ times the probability  when  not  having $P^*$. For a fixed $\epsilon$, the larger $\gamma$ is, the larger the privacy loss $(\epsilon\gamma)$ will be.  For example, Tom is standing in Times Square in NYC and looking for a restaurant for lunch. He sends a request to a service provider for a list of restaurants nearby. However, he does not want to disclose his exact location and chooses to release a perturbed location $P^*$ via GI with $\epsilon=0.1$ per mile. The probability the service provider will identify his true location within a radius of $\gamma=1$ mile given the perturbed location information is at most 1.1 folds of the probability when  not having the information, and at most 148 folds within a radius of $\gamma=50$ miles. In the latter case, though the probability of distinguishing locations given $P^*$ dramatically increases compared to not having $P^*$, it is not practically alarming from a privacy perspective as the increase is caused by the large $\gamma$ rather than a large $\epsilon$. In other words, the service provider will have great confidence that Tom is in NYC given the released $P^*$, but little confidence in pinpointing exactly where in NYC. If it were the combination of $\epsilon=50$ and $\gamma=1$, then the probability the service provider identifying Tom's true location within a radius of 1 mile would increase by 248 folds  given the perturbed location information, constituting a disastrous situation in privacy.

The planar Laplace mechanism can be used to achieve $\epsilon$-GI by perturbing location information in polar coordinates.
\begin{defn}[\textbf{polar Laplace mechanism} \cite{andres2013geo}]\label{def:polar}
The sanitized location $P^*$,  given the actual location $P$ with coordinates $(x,y)$ in the Euclidean space, satisfies GI with coordinates
\begin{equation}\label{Eq:xsys}
(x^*,y^*)=(x+r\cos(\theta), y+r\sin(\theta)), 
\end{equation}
where  the joint distribution of $R$ and $\theta$ is
\begin{equation}\label{Eq:planar}
f(r,\theta)=\epsilon^{2}re^{-\epsilon r}/(2\pi).
\end{equation}
\end{defn}\vspace{-6pt}
Eq (\ref{Eq:planar}) implies $R$ and $\theta$ are independently distributed and 
\begin{align}
r&\sim\mbox{gamma}(2,\epsilon)=r\epsilon^2 e^{-\epsilon r}\label{Eq:r}\\
\theta&\sim\mbox{Unif}(0,2\pi)=1/(2\pi).\label{Eq:theta}
\end{align}
In summary, to generate a sanitized location $P^*$, one may draw $R$ from the gamma distribution with shape 2 and scale $\epsilon^{-1}$ and $\theta$ from Unif($0,2\pi$), and then calculate the coordinates of $P^*$ in the Euclidean space per Eq (\ref{Eq:xsys}).

\vspace{-9pt}\section{Privacy-preserving TPU with GI}\label{sec:method}\vspace{-3pt}
Applied to the collection of GPS records, the GI notion can help protect individual privacy on several types of information: an individual location at a given time point, the travel trajectory of an individual over a time period, and any derived information from the collected sanitized trajectories, such as TPU. 

In what follows, we present a procedure to achieve PP-TPU in the framework of GI, taking into the composability of privacy costs from disclosing multiple location points from a trajectory and leveraging public knowledge of maps and road networks to improve the utility of PP-TPU on a given target road. We also examine the accuracy of sanitized information relative to the original  information; analyze the privacy guarantees of the proposed procedure, along with newly proposed metrics for quantifying adversary errors. 

\vspace{-9pt}
\subsection{Proposed PP-TPU Procedure}\label{subsec:method}\vspace{-3pt}
We propose  a new PP-TPU procedure. The PP-TPU procedure sanitizes GPS records locally via the planar Laplace mechanism to guarantee GI before the information is shared with the service provider (the strategy in Fig \ref{fig:adversary}(c)). This approach mitigates the privacy risks of learning new private information about an individual from the collected GPS records for various types of adversaries,  as only the users themselves possess the true responses. We also take several measures to improve the accuracy of the proposed PP-TPU procedure and to quantify  the utility of sanitized trajectories, detailed below.

First, given a fixed total per-trajectory privacy cost, we limit the number of records to be collected per traveler so that the sanitization of each location record does not inject too much noise to render the sanitized record useless.

Second, we filter out  non-usable trajectories for the PP-TPU  given a target route $R$. Due to the sanitization noise injected to satisfy GI at each location, the travel direction between two consecutive time points may be opposite to route $R$, which has  a pre-specified direction. Not to bias the total travel distance, we keep the  sanitized locations as is, as long as they can be mapped onto route $R$, but attach a sign to indicate the travel direction consistency with route $R$, namely, positive distance if the traveling direction  is consistent with the direction of route $R$, negative if opposite, and 0 if the two mapped locations completely overlap. After the complete set of the GPS records from the traveler is mapped, we sum the signed distances on $R$ for the traveler. If the summed distance is negative, then the trajectory is not usable, as defined in Definition \ref{def:usable}.
\begin{defn}[usable trajectory]\label{def:usable} A usable trajectory given a target route is a trajectory that satisfies the following two conditions: 1) at least two consecutive locations are mapped onto the target route $R$; 2) the total travel distance summed over distance segments calculated from the mapped coordinates on $R$ is non-negative.  The set of usable trajectories is the \emph{usable set $\mathcal{U}$}.
\end{defn}
\vspace{-3pt}
Third, we provide users an option to weigh different trajectories for their various levels of contribution towards the TPU on a given target route $R$. The motivation behind this is as follows. It is very likely that not all the GPS records will be mapped to route $R$, even if the traveler stays on $R$ all the time at least for the period of interest, for a few reasons. First,  GPS information is not always accurate due to satellite signal blockage and multipath effects, causing difficulty in road matching. Second, road mapping procedures themselves often involve approximation and errors. Third, with the additional randomness introduced by the GI sanitization, the location accuracy will further decrease. Therefore, each trajectory may have a different number of GPS records mapped onto $R$, some of which are consecutive in times and others are not. When calculating the travel distance on $R$ for a traveler, it makes sense to only count the distances between the locations at two consecutive time points if both are mapped onto  $R$. One way to formulate the weight is to let it be proportional to how much a sanitized mapped trajectory overlaps with the target route.
\begin{defn}[trajectory weight]\label{def:weight} Denote by $d^*_i$ the travel distance for traveler $i$ on the target route $R$ of length $d$ from the usable set $\mathcal{U}$. The weight that traveler $i$ carries in the TPU is  $w_i=d^*_i/d$.
\end{defn} \vspace{-3pt}

Algorithm \ref{alg:mechanism} lists the steps of our proposed PP-TPU procedure, with the above three utility-improvement measures implemented in various stages of the procedure.  \vspace{-6pt}
\begin{algorithm}
\caption{The PP-TPU Procedure}\label{alg:mechanism}
\SetAlgoLined
\SetKwInOut{Input}{input}
\SetKwInOut{Output}{output}
\Input{GPS location coordinates $(x_{ij},y_{ij})$ with timestamp $\tau_{ij}$ for $i\!=\! 1,\ldots,K$ trajectories and $j\!=\!1,\ldots, n_i (\le n$ the maximum records per trajectory); per-trajectory privacy budget $\epsilon_i$; target route with total distance $d$.} 
\Output{sanitized travel time $\mathbf{t}^*$, trajectory weight $\mathbf{w}$.} 
Usable set $\mathcal{U}\leftarrow\emptyset$\; 
\For{$i=1,\ldots,K$}{
$d^*_i\leftarrow0$; $\delta t_i\leftarrow0$\;
\For{$j=1,\ldots, n_i$}{
Perturb $P_j=(x_{ij},y_{ij})$ via the planar Laplace mechanism in Eq (\ref{Eq:xsys}) with privacy budget $\epsilon_i/n_i$ to yield $P^*_j=(x^*_{ij},y^*_{ij})$ at time $\tau_{ij}$\;
Map $P^*_j$ onto the area map to obtain the map coordinates $Q^*_j$\;
\If{($Q^*_{j-1}, Q^*_j$) for $j>1$ fall on Route $R$}{
Calculate the signed Euclidean distance $d_{ij}$ between  $Q^*_{j-1}$ and $Q^*_j$\;
$d^*_i\!\leftarrow\! d^*_i+d^*_{ij}$; $\delta t_i\!\leftarrow\! \delta t_i+(\tau_{ij}-\tau_{i,j-1})$\;} 
}
\If{$d^*_i\ge0$}{
$\mathcal{U}\leftarrow\mathcal{U}\cup i$\;
Calculate speed $s^*_i\!\!=\!d^*_i/(\delta t_i)$, predicted travel time $t^*_i=d/s^*_i$, and weight $w_i=d^*_i/d$.}
}
\end{algorithm} 
\vspace{-3pt}
With the output weights $\mathbf{w}$ from Algorithm \ref{alg:mechanism}, we can calculate the effective number of mapped full trajectories to provide an overall metric on the impact of mapping  and sanitation of GPS records on the TPU on a target road.
\begin{defn}[effective number of mapped full trajectories]\label{def:eff} The effective  number of mapped  full trajectories is $K_{\text{eff}}=\sum_{i\in\mathcal{U}}w_i$.
\end{defn}
Since $w_i\in[0,1]$, $K_{\text{eff}}\le|\mathcal{U}|$, where $|\mathcal{U}|$ is the number of trajectories in  $\mathcal{U}$. $|\mathcal{U}|$ in turn is $\le K$, where $K$ is the number of raw GPS trajectories before mapping, sanitation, and filtering out. $K_{\text{eff}}$ in a PP-TPU depends on $\epsilon$, the number of  GPS trajectories $K$ before mapping, and the pattern and complexity of the road networks onto which the GPS records are projected.   Besides using weights to calculate $K_{\text{eff}}$, we can also incorporate the weights in the TPU by define a weighted version of $f^*_w(t)$. For example, we may sample $K_{\text{eff}}$ travel times from set ($t^*_1,\ldots,t^*_{|\mathcal{U}|}$) with the sampling probabilities proportional to $\mathbf{w}=\{w_1,\ldots,w_{|\mathcal{U}|}\}$ and obtain an empirical $\hat{f}^*_w(t)$ based on the samples.

\vspace{-6pt}\subsection{Accuracy of Sanitized Information}\label{sec:accuracy}\vspace{-3pt}
As mentioned above, road mapping procedures per se involve approximation and errors, the quantification of which is challenging and case-dependent. As such, we focus on the accuracy of the perturbed GPS records relative to their original, instead of on the mapped coordinates. It is reasonable to assume that if sanitized and original GPS records are close, so are their mapped locations. 

\begin{figure*}[t]
\includegraphics[width=1\textwidth, height=0.18\textheight]{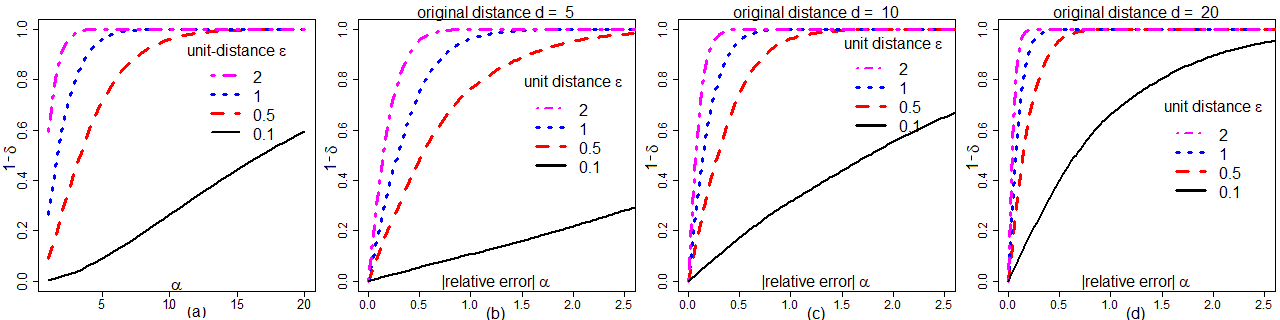}\vspace{-3pt}
\caption{Usefulness analysis on perturbed GPS location (a) and distances (b) to (d)} \label{fig:useful}\vspace{-9pt}
\end{figure*}

We quantify  the closeness between a  sanitized GPS location vs its original using the ``usefulness'' definition \cite{andres2013geo}. A location perturbation mechanism is $(\alpha, \delta)$-usefulness if the distance between the sanitized  and original locations is $\le\alpha$ with a probability of $1-\delta$, for every original location. For example, for a unit-distance privacy budget $\epsilon=2$, the probability that a sanitized location via the planar Laplace mechanism is within $\alpha=1.5$ units of the original location is $1-\delta=0.8$, calculated directly from the CDF of gamma(2, 1.5). In other words, the planar Laplace mechanism  of $\epsilon=2$ GI is $(1.5,0.2)$-useful for sanitizing locations. We plot the relationships between $\alpha$ and $1-\delta$ for a range of $\epsilon$ values for the planar Laplace mechanism  in Fig. \ref{fig:useful}(a). 

In addition, we may assess the accuracy of the distance between two  sanitized  locations.  Denote by $(x_j,y_j)$ and  $(x_{j'},y_{j'})$ the coordinates  of two recorded GPS locations at times $\tau_j$ and $\tau_{j'}$, respectively. The sanitized coordinates for the two locations via the planar Laplace mechanism in Eq (\ref{Eq:planar})  are respectively, 
\begin{align}
\begin{cases}
x^*_j=x_j+r\cos(\theta),\; y^*_j=y_j+r\sin(\theta)\\
x^*_{j'}=x_{j'}+r'\cos(\theta'),\; y^*_{j'}=y_{j'}+r'\sin(\theta')
\end{cases},
\end{align}
the distance between which can be calculated by the Euclidean distance 
\begin{align}
&d^{*2}_{jj'} = (x^*_j\!-x^*_{j'})^2 +(y^*_j\!-y^*_{j'})^2 = d^2_{jj'} +\Delta_{jj'},\mbox{ where}\label{Eq:dgeneral}\\
&\Delta_{jj'}\!=\!r^2\!+\!r'^2\!-\! 2rr'(\cos(\theta_j)\cos(\theta_{j'})+\sin(\theta)\sin(\theta'))+\notag\\ 
&\qquad\quad2(x_j-x_{j'})(r'\cos(\theta')-r\cos(\theta))+\notag\\ 
&\qquad\quad2(y_j-y_{j'})(r'\sin(\theta')-r\sin(\theta)),\label{Eq:delta}
\end{align}
and $d_{jj'}$ is the Euclidean distance between the original GPS records at times $\tau_j$ and $\tau_{j'}$. $\Delta_{jj'}$ can be regarded as the bias of the squared sanitized distance from the original distance,  $d^*_{jj'}$ conditional on  $d_{jj'}$  is a random variable as $r,r',\theta,\theta'$ are all random variables. We propose two metrics to examine the accuracy of $d^*_{jj'}$ relative to $d_{jj'}$. 

For the first metric,  we define $(d,\alpha,\delta)$-usefulness for sanitized distances, in a similar manner   to the $(\alpha,\delta)$-usefulness in general \cite{blum} and for sanitized locations \cite{andres2013geo}.
\begin{defn}[\textbf{$(d,\alpha,\delta)$-usefulness of sanitized distance}] A randomization mechanism is $(d,\alpha,\delta)$-useful, if there is a probability of $1-\delta$ that  the sanitized distance $d^*$ satisfies $|d^*\!/d-1|<\alpha$ for every pair of locations with a distance of at least $d$.
\end{defn}\vspace{-3pt}
$\alpha$ is the relative error of the sanitized $d^*$ to the original $d$.  The smaller $\alpha$ and the larger $\delta$ are for a given $d$, the more useful the mechanism is in terms of distance preservation. Figs. \ref{fig:useful}(b) to \ref{fig:useful}(d) depict the relationship  between $\alpha$ and $\delta$ when the original distance $d$ is 5, 10, and 20 at different levels of unit-distance privacy cost $\epsilon$. As $d$ increases, $\delta$ decreases for the same $\alpha$. From the plots, we can claim that there is a 80\% probability that the distance $d^*$ between the perturbed locations via the planar Laplace mechanism of $\epsilon=1$ GI is within $\pm 25\%$ of $d\ge10$; in other words, the mechanism is $(10,  0.25, 0.2)$-useful at $\epsilon=1$. Similarly, we may also claim the mechanism is  $(5, 0.5, 0.2)$-useful for $\epsilon=1$, and $(5, 1.0, 0.3)$-useful for $\epsilon=0.5$, etc.

For the second utility  metric on sanitized distances, we calculate the expected \%deviation  $\E(d^*_{jj'}/d_{jj'}\!-\!1)$  and the \% root mean squared deviation (\%RMSD) $\sqrt{\E(d^*_{ij}/d_{ij}\!-\!1)^2}$ of the  sanitized distance from the original distance, respectively. Eqs (\ref{Eq:dgeneral}) and (\ref{Eq:delta}) suggest there is no closed-form expression for either of them; but we can always examine the numerical deviations for a given scenario. Table \ref{tab:bias} lists the expected \%deviation and \%RMSD in distance for different scenarios of $\epsilon$ and $d$.  As expected, the larger $\epsilon$ or the larger $d$ is, the smaller the \%deviation is.  Also listed in the table is the  expected \%deviation in squared distance  $\E(d^{*2})/d^2)\!-\!1$, which has a closed-form solution.  Specifically, $\E(r^2)=\E^2(r)+\V(r)=6\epsilon^{-2}$ ($\V$ denotes variance), so is $\E(r'^2)$; since $2(x_j-x_{j'})(\E(r')\E(\cos(\theta'))-\E(r)\E(\cos(\theta))+2(y_j\!-\!y_{j'})(\E(r')\E(\sin(\theta'))\!-\!\E(r)\E(\sin(\theta))=0$, then  
\begin{align}
\E(d^{*2}_{jj'}-d^2_{jj'})&= 12\epsilon^{-2}=O(\epsilon^{-2});\label{eqn:deviance}\\
\mbox{and } \E(d^{*2}_{jj'}/d^2_{jj'}-1)&= 12/(d_{jj'}\epsilon)^{2}. \label{eqn:pdeviance}
\end{align}
Eq (\ref{eqn:deviance}) indicates that, in expectation, the squared distance between two sanitized GPS locations always deviates from the squared original distance by a constant $12\epsilon^{-2}$ for a given $\epsilon$, regardless of $d_{jj'}$; however, Eq (\ref{eqn:pdeviance}) implies that the deviation is not meaningful for large $d_{jj'}$.

\vspace{-9pt}\subsection{Privacy Guarantee and Adversary Error} \label{sec:AE}\vspace{-3pt}
As illustrated in Fig \ref{fig:adversary}(c), the proposed PP-TPU procedure is based on sanitized GPS trajectory data, mitigating the privacy risk from both the honest-and-curious and malicious adversaries. The employed privacy model, GI, is an extension of the notion of DP to location settings with a
\begin{table}[!htb]
\caption{Expected \%deviation and \%RMSD in distance, and expected \%deviation in squared distance}
\label{tab:bias}\vspace{-6pt}
\centering
\resizebox{!}{0.125\columnwidth}{
\begin{tabular}{c@{\hspace{0.3\tabcolsep}}|@{\hspace{0.5\tabcolsep}} c@{\hspace{0.75\tabcolsep}}c@{\hspace{0.75\tabcolsep}}c@{\hspace{0.75\tabcolsep}}|@{\hspace{0.5\tabcolsep}} 
c@{\hspace{1\tabcolsep}}c@{\hspace{0.5\tabcolsep}}c|@{\hspace{0.5\tabcolsep}}
c@{\hspace{1\tabcolsep}}c@{\hspace{1\tabcolsep}}c@{\hspace{0.5\tabcolsep}} }
\hline
& 
\multicolumn{3}{@{\hspace{0.1\tabcolsep}}c@{\hspace{0.2\tabcolsep}}|}{$\left(\frac{\E(d^{*})}{d}\!-\!1\!\right)(\%)^\dagger$}&
\multicolumn{3}{@{\hspace{0.1\tabcolsep}}c@{\hspace{0.2\tabcolsep}}|}{$\sqrt{\E(\frac{d^*}{d}\!-\!1)^2}(\%)^\dagger$}&
\multicolumn{3}{c}{$\left(\frac{\E(d^{*2})}{d^2}\!-\!1\!\right) (\%)^\ddagger$} \\
\cline{1-10}
$d$  & 50 & 100 & 200 & 50 & 100 & 200 & 50 & 100 & 200 \\
\hline
$\epsilon\!=\!0.01$ & 5.00 & 2.09 & 0.75 & 6.17 & 2.80 & 1.23 & 48 & 12 & 3\\
$\epsilon\!=\!0.05$ & 0.51 & 0.13 & 0.03 & 0.95 & 0.47 & 0.24 & 1.92 & 0.48 & 0.12\\
$\epsilon\!=\!0.25$ & 0.02 & 0.00 & 0.00 & 0.19 & 0.10 & 0.05 &  0.0768 & 0.0192 & 0.0048 \\
\hline
\end{tabular}}
\begin{tabular}{l}
\footnotesize $^\dagger$ numerical results; $^\ddagger$ analytical results via  Eq. (\ref{eqn:pdeviance}). \hspace{0.75in}\textcolor{white}{.}\\
\hline
\end{tabular}\vspace{-12pt}
\end{table}
similar mathematical concept for controlling privacy loss when sharing information. DP is known to provide "provable privacy protection against a wide range of potential attacks, including those currently unforeseen'' \cite{nissim2017differential,dwork2017exposed}.  The proposed PP-TPU procedure in Sec \ref{subsec:method} protects several types of spatial-temporal information: the location of a traveler at a given time point, a travel trajectory of the traveler for a given time period, any calculated statistics from the trajectory (e.g, travel distance, travel speed) per the immunity property of DP and GI against post-processing. We examine each yielded privacy protection type below in detail, especially in the case  of a travel trajectory.

First, per the definition of GI in Definition \ref{def:GI}, the probability of distinguishing the true location $P$ from any other locations with a radius of $\gamma$, given the released perturbed location $P^*$ increases by $e^{\epsilon\gamma}-1$ folds compared to the probability  when not having $P^*$. In other words, the same privacy guarantees and indistinguishability as illustrated in Definition \ref{def:GI}  apply to the GPS records collected at each timestamp for the PP-TPU. 

Second, the proposed PP-TPU procedure protects the privacy of a collected travel trajectory over a time period. Though each of the location records  on the trajectory is perturbed via the planar Laplace mechanism has a straightforward interpretation on indistinguishability as presented above, how to quantify the adversary error in the learning of the original trajectory based on the released sanitized trajectory is less studied. Below we propose two metrics --  the \emph{ average distance (AD)} and  the \emph{consecutive positioning degree (CPD)} -- to quantify the adversary error and assess the effectiveness of a randomization procedure in protecting travel trajectory privacy.  We apply both metrics to examine the adversary error in the experiments in Sec. \ref{sec:exp}.
\begin{defn}[average distance]\label{def:etd}
The average distance (AD) between the sanitized and  original mapped travel trajectories on a  road network is the averaged distance between the two sets of mapped locations at the same set of timestamps from the two trajectories.
\end{defn}
We may calculate the AD empirically as follows. The pair of original and sanitized coordinates of the mapped trajectories $i$ ($i=1,\ldots,K$) are $\{(x_{ij},y_{ij})\}$  and  $\{(x^*_{ij},y^*_{ij})\}$, respectively, at time $\{\tau_{ij}\}$ for $j=1,\ldots,n_i$. The AD is given by
\begin{equation}\label{eqn:AD}
\textstyle K^{-1}\sum_{i=1}^K\left(n_i^{-1}\sum_{j=1}^{n_i}  d((x_{ij},y_{ij}), (x^*_{ij},y^*_{ij}))\right).
\end{equation}
Given a set $\{n_i\}_{i=1,\ldots,K}$, the larger AD, the larger the adversary error and the more difficult it is to recover the original trajectory from the sanitized trajectory (the reason that we define AD rather than ``total distance'' -- the summed distances between the GPS locations from two trajectories is that AD corrects for $n_i$, which may differ by trajectory).

\begin{defn}[consecutive positioning degree]\label{def:CPD}
The consecutive positioning degree (CPD) $p(l)$ is a probability  distribution  of correctly identified $l$ consecutive locations on a trajectory based on the released sanitized trajectory with $n$ GPS records, for $l=0,\ldots,n$. The expected value of correctly identified positions out of $n$ is $n_c=\sum_{l=0}^n l\times p(l)$.
\end{defn}

We choose to examine $p(l)$, the  distribution of correctly identified  \emph{consecutive} positions instead of correctly identified positions $p(m)$ for $m=0,\ldots,n$ (regardless of whether they are consecutive or not) because the former  would be regarded by many as more revealing of travel trajectory and carrying more privacy concern than latter. How to define ``correctly identified positions'' is up to the user.  One approach is hard-thresholding. Specifically, we choose a clip radius $C$. If the sanitized location falls within the circle of radius $C$  centered at the original location, then it is claimed as correct positioning. The smaller $C$ is, the harder it is to meet the criterion, but the more meaningful ``correct'' is. Since each location on a trajectory is perturbed independently via the polar Laplace mechanism, with the hard-thresholding rule, the probability of correctly identifying a location can be determined analytically, which is $p=F(C;2,\epsilon/n)$, where $n$ is the number of recorded positions on a GPS trajectory and $F$ is the CDF of gamma($2,\epsilon/n$). 

The number of correctly identified positions $m$ given $p$ follows $m\sim$ Binomial$(n,p)$. As for the distribution of CPD $l$, we can leverage Binomial$(n,p)$ to express $p(l)$ analytically when $n$ is small, but $p(l)$ for $1\le l< n-k$ with small $k\ge0$ becomes less tractable as $n$ increases considering that a trajectory may contain multiple location strings of different $l$. For example, a GPS trace with $n=10$ records may have 2 occurrences of $l=1$, 1  occurrence of $l=2$, and 1  occurrence of $l=3$. For cases where analytical calculation of $p(l)$ becomes difficult, we design Algorithm \ref{alg:CPD} that uses Monte Carlo (MC) simulations to calculate $p(l)$. Though the algorithm is presented with the hard-thresholding rule for correct positioning (line 4), the steps are applicable to other definitions of correct positioning. $n_i^{(l)}$ in the algorithm  refers to the frequency distribution $l$ in trajectory $i$, its average over $K$ trajectories gives the MC estimate $p(l)$. The algorithm also outputs $\bar{n}_c$, the MC estimate of the expected  value  of  correctly identified positions $n_c$ in Definition \ref{def:CPD}.

\vspace{-9pt}\section{Experiments}\label{sec:exp}\vspace{-3pt}
We conduct four experiments to investigate empirically the impact of sanitization of GPS trajectories on the utility of TPU in four road network scenarios. In each experiment, there is a pre-specified target route on which the TPU 
\setlength{\textfloatsep}{0pt}
\begin{algorithm}
\caption{Calculation of CPD $p(l)$}\label{alg:CPD}
\SetAlgoLined
\SetKwInOut{Input}{input}
\SetKwInOut{Output}{output}
\Input{$K$ GPS trajectories and their sanitized counterparts with $n$ records per trajectory; clip radius $C$} 
\Output{$n^{(l)}_i$ for $i=1,\ldots,K$; $p(l)=\sum_{i=1}^K n^{(l)}_i\left(\sum_{l=0}^n\sum_{i=1}^K n^{(l)}_i\right)^{-1}$;  $\bar{n}_c=K^{-1}\sum_{i=1}^K \sum_{l=0}^n(n_i^{(l)}\times l)$.}
\For{$i=1,\ldots,K$}{
\For{$j=1,\ldots,n$}{
Calculate the distance $d_{ij}$ between sanitized location $P^*_{ij}$ and original location $P_{ij}$\;
If $d_{ij}\le C$, then $e_{ij}=1$; else $e_{ij}=0$\; 
Let $e_{i0}=0$ and $e_{i,n+1}=0$\;}
If $e_{ij'}\!=\!0\;\forall j'\!=\!1,\ldots,n$, then  $n^{(0)}_i=1$; else $n^{(0)}_i=0$\;
If $e_{ij'}\!=\!1\;\forall j'\!=\!1,\ldots,n$, then  $n^{(n)}_i=1$; else $n^{(n)}_i=0$\;
\For{$l=1,\ldots,n-1$}{
$n^{(l)}_i\leftarrow0$\;
\For{$j=1,\ldots,n-l+1$}{
\If{$(e_{ij'}\!=\!1\;\forall j'\!=\!j,\ldots,j\!+\!l\!-\!1)
\;\&\; (e_{i,j-1} =0) \;\&\; (e_{i,j+l} =0$)}
{$n^{(l)}_i \leftarrow n^{(l)}_i+1$}}
}}
\end{algorithm} 
analysis is performed.  We examine the utility of PP-TPU for a range $\epsilon$ values and assess the adversary error in learning individual trajectories. Though a privacy-preserving travel time distribution may also be obtained by sanitizing the original empirical distribution via a DP mechanism, as illustrated in Fig.~\ref{fig:adversary}(b), the server needs to collect the actual individual GPS locations, and the sanitization is processed on the server. Therefore, this approach does not provide the same privacy guarantees as the decentralized and local approach (Fig.~\ref{fig:adversary}(c)) taken by Algorithm \ref{alg:mechanism}. Since it is impossible to match the level of privacy protection between the two approaches, the utility comparison would not be as meaningful; therefore, we choose not to compare our PP-TPU approach with the DP-based approach in the experiments.

\vspace{-9pt}\subsection{Experiment Settings}\label{sec:setting}
\begin{figure}[!htb]
\includegraphics[height=0.2\textwidth, angle=90]{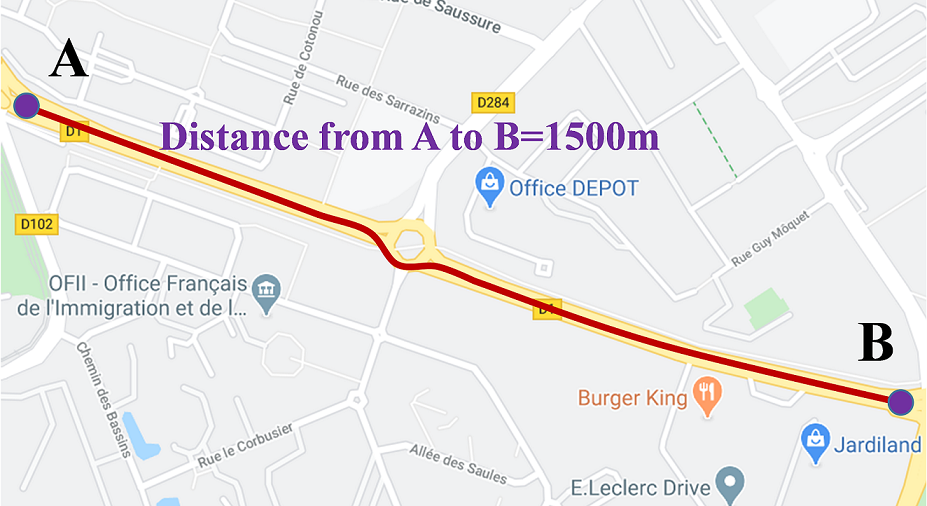}
\includegraphics[width=0.27\textwidth]{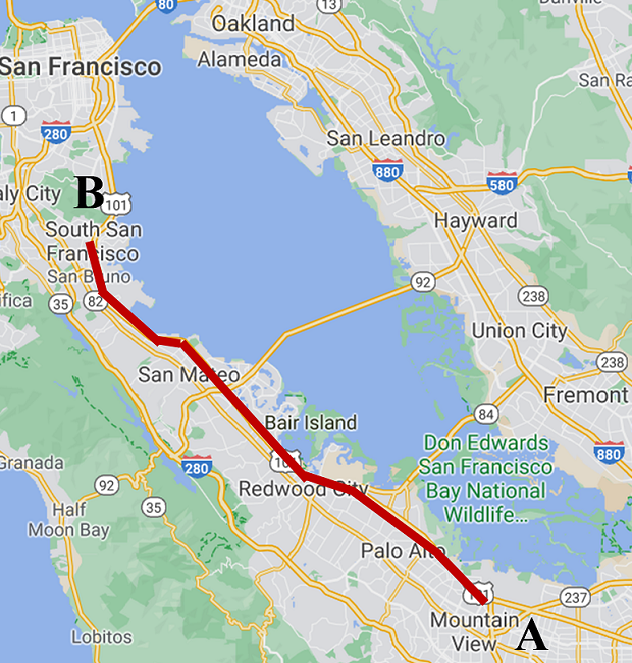}\\
\centering (a) experiment 3 \hspace{0.7in} (b) experiment 4 \\
\centering\hspace{0.1in} Creteil, France  \hspace{0.7in}  San Francisco, USA\\
\vspace{-6pt}
\caption{Area maps in Experiments 3 and 4 (source: Google Map). AB is the target route for TPU in each experiment.} \label{cologne}
\end{figure}

\vspace{-3pt}
In Experiment 1, the simulated  road network contains a single road. In Experiments 2, the simulated  road network contains three parallel roads with one being the target route.  In Experiment 3, the road network is around a large  roundabout in the town of Creteil in France (Fig \ref{cologne}(a)); the target route AB is about 1.5 kilometers long. In Experiment 4, we examine a region in the San Francisco Bay Area (Fig \ref{cologne}(b)); the target road AB is about 50 kilometers long. 

The GPS trajectory data in Experiments 1 (1,000 trips) and 2 (1,000 trips on the target road) are simulated as follows. We first simulated speeds from the inverse Weibull distribution with mean $\mu=24$ meter per second and variance $\sigma^2=8$ (the values are chosen to mimic some common real traffic speed distributions). Each simulated speed corresponds to one trip, on which 10 location records were generated at a fixed timestamp of every $\tau=20$ seconds, leading to travel trajectories of different lengths, depending on the speed. The vehicular mobility trace data in Experiment 3 \cite{vehicular} are  downloadable from \url{http://vehicular-mobility-trace.github.io/} and  contains 857,136  sets of location coordinates per second from around 5102 trips during the morning rush hour (7 to 9 AM), simulated based on real data. We randomly chose 1,000 trips within the rectangle bounded by the coordinates of the ends points A and B of the target route. The dataset in Experiment 4 \cite{sanf} contains real  mobility traces of taxi cabs and is downloadable from \url{http://crawdad.org/epfl/mobility/20090224/index.html}. It contains the GPS coordinates of approximately 500 taxis  over 30 days. For this experiment, we used a subset of 30,900 location-time GPS records over the morning  rush  hours (8 to 9 AM) from  419 trips.  In Experiments 3 and 4, we set the maximum number of GPS records per trip at 10 so to control the privacy loss per traveler. If a traveler has $\le10$ records, we used all of them; otherwise, we randomly sampled 10 records or  had  10 records spaced equally over the trajectory if there were enough records to allow that.

\vspace{-9pt}\subsection{Sanitization and Implementation Details}\vspace{-3pt}
The GPS records were sanitized via the  planar Laplace mechanism and projected into the road map in each experiment using the shortest path algorithm. The PP-TPU was then conducted via algorithm \ref{alg:mechanism} in each experiment. For the GI sanitization, we set the per-location per-meter privacy loss at 0.005, 0.01, 0.03, 0.05, and 0.08 in all 4 experiments. Since the maximum of GPS records per trip is 10, the total privacy cost for releasing a trajectory is $\le 0.05, 0.1, 0.3, 0.5, 0.8$, respectively.   

\begin{figure}[!htb]
\vspace{-6pt}\centering
 Experiment 2 \\
\includegraphics[width=0.48\textwidth, height=0.23\textwidth]{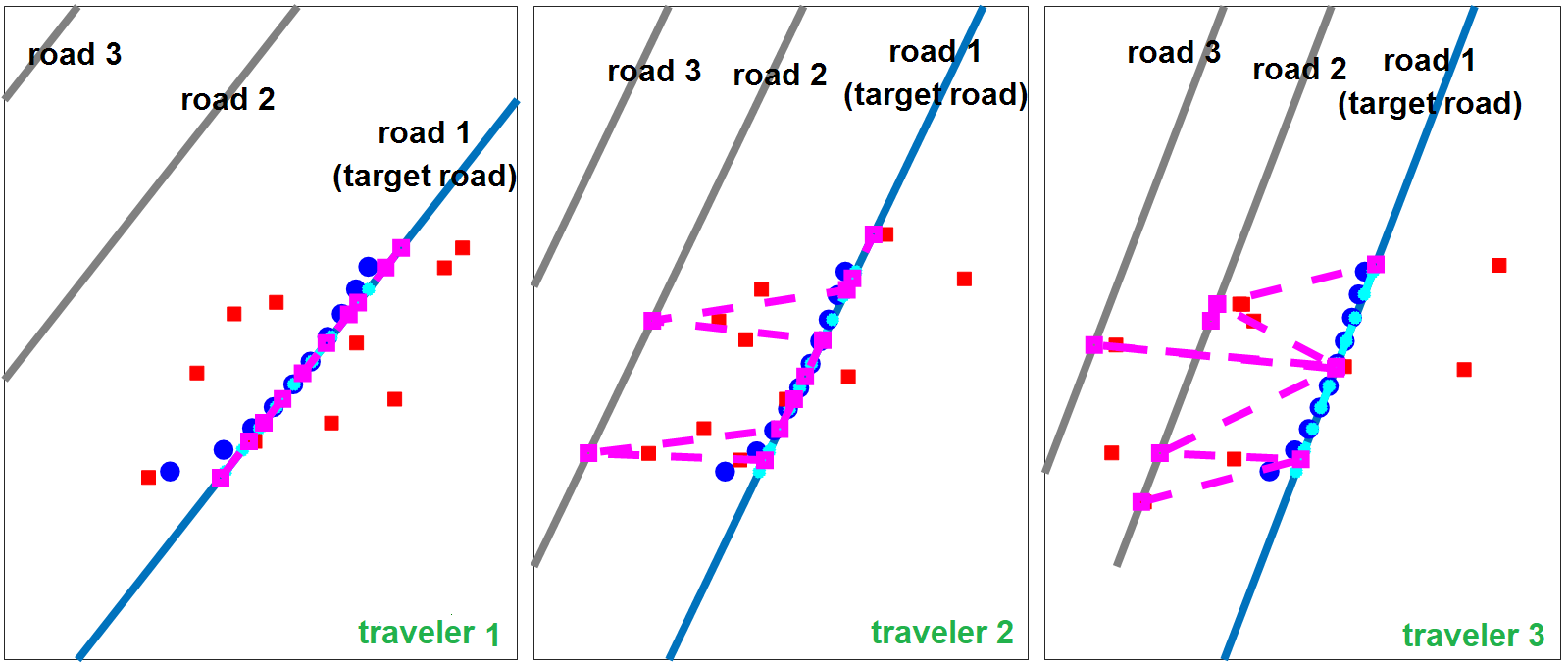}\\
\vspace{3pt} 
Experiment 1 \hspace{0.3in} Experiment 3 \hspace{0.3in} Experiment 4\\
\includegraphics[width=0.48\textwidth, height=0.22\textwidth]{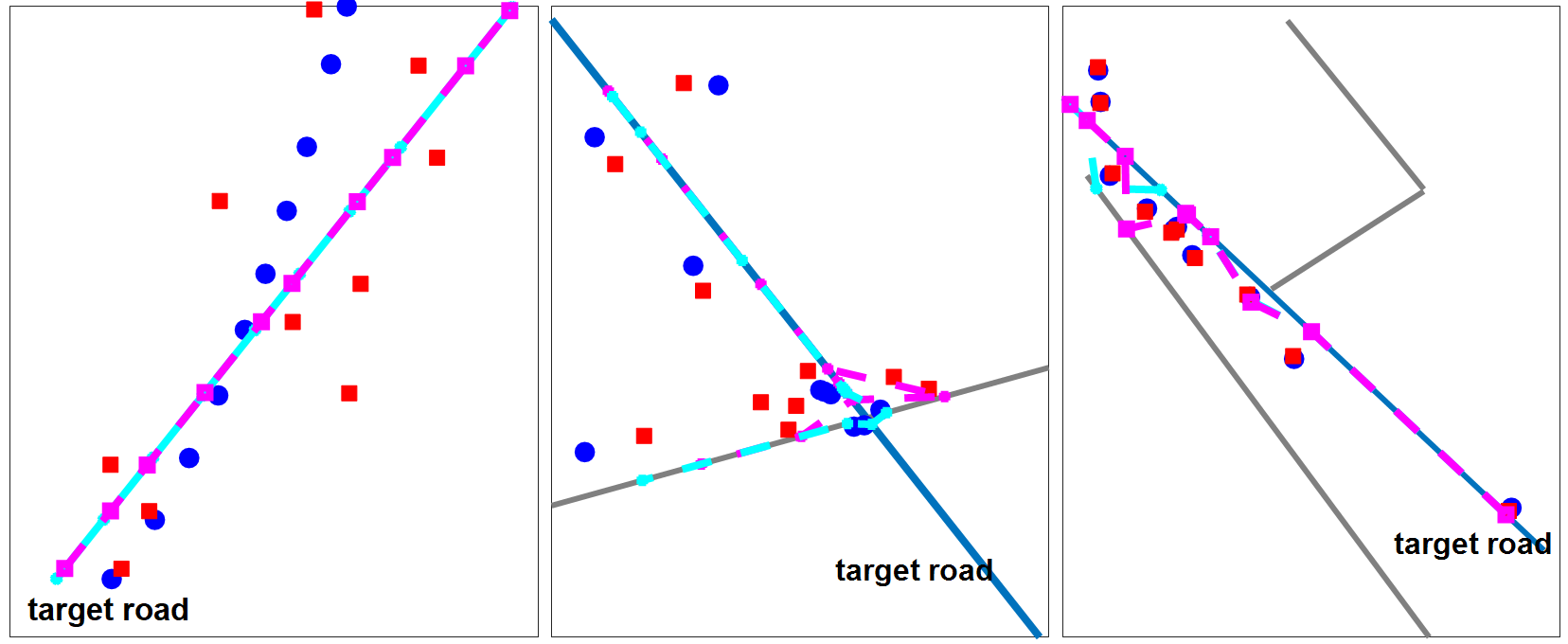}\\
\includegraphics[width=0.45\textwidth]{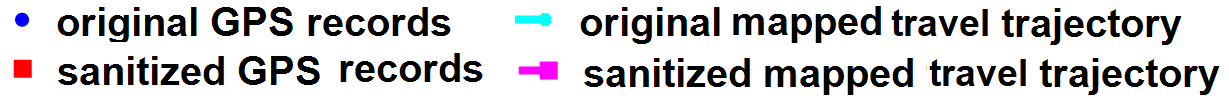}
\vspace{-6pt}
\caption{Examples of  sanitized GPS records and mapped travel trajectories at per-trajectory privacy cost of $\epsilon=0.1$}\label{fig:road}
\vspace{-6pt}
\end{figure}

Fig \ref{fig:road} presents some examples of  sanitized GPS records and mapped travel trajectory on road networks given the GPS records. Take Experiment 2 as an example. Road 1 is the target road for TPU analysis. If there was no privacy concern, the three travelers would share their GPS records (blue circles) with the service provider who would project the records via a mapping algorithm onto the road network and use usable travel trajectories on road 1 to calculate travel time and carry out TPU. In this case, the mapped trajectories (cyan lines) fall on the target road for all three travelers. For PP-TPU, the service provider collect only sanitized versions (red squares) of the original GPS records; the mapping procedure and TPU analysis are the same as in the non-private setting. Since the sanitized GPS records deviate from their original counterparts, it is almost certain the trajectories after mapping also deviate from the original. For traveler 1, all ten sanitized GPS records are mapped onto road 1 and can be used for the subsequent travel time calculation. For traveler 2, eight out of the ten  sanitized GPS records are mapped on road 1 and two on the nearby road 2.  The eight records on road 1 form two location strings of length $l=4$ and $l=2$, respectively, that are used for the subsequent PP-TPU analysis. For traveler 3, three out of the ten sanitized GPS records are mapped onto road 1 but none of the two are consecutive in time, so traveler 3 does not contribute toward the PP-TPU. In summary, out of the sanitized trajectories from the three travelers, only those from travelers 1 and 2 contribute to $\mathcal{U}$.

\vspace{-16pt} \subsection{Utility and PP-TPU Results}\label{sec:utility}
\begin{figure}[!htb]
\centering
\hspace{0.3in} Experiment 1 \hspace{0.8in} Experiment 2 \\
\includegraphics[width=0.245\textwidth]{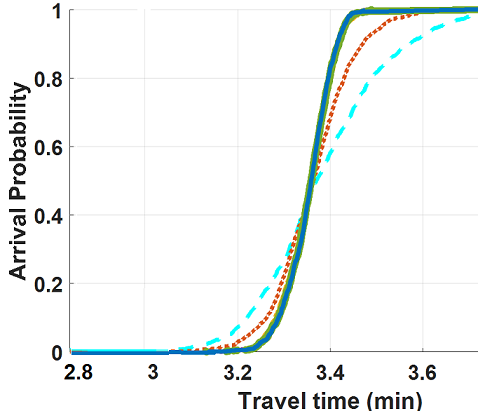}\hspace{-4pt}
\includegraphics[width=0.245\textwidth]{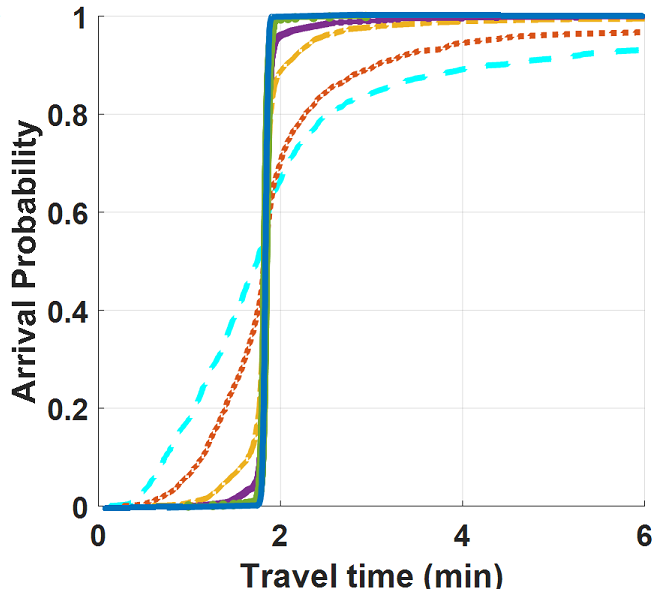}\hspace{-4pt}
\hspace{0.3in} Experiment 3 \hspace{0.8in}  Experiment 4\\
\includegraphics[width=0.49\textwidth]{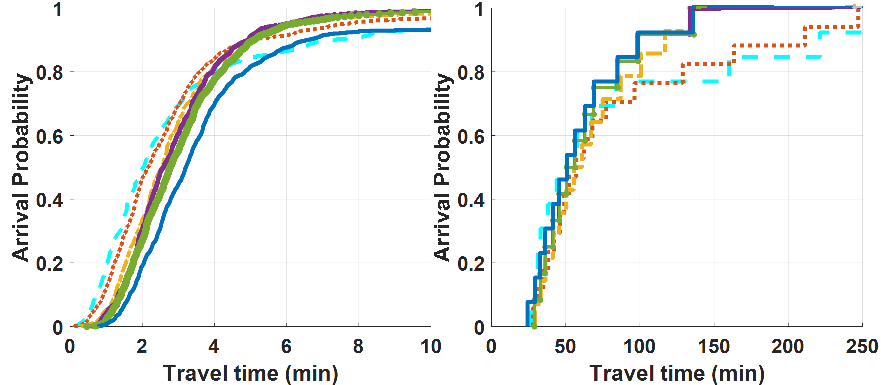}
\includegraphics[width=0.4\textwidth]{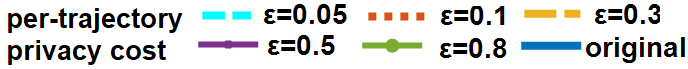}
\vspace{-6pt}\caption{PP-TPU}\label{fig:CDF}\vspace{-9pt}
\end{figure}

\begin{figure}[!htb]
\centering
\hspace{0.3in} Experiment 1 \hspace{0.8in}  Experiment 2\\
\includegraphics[width=0.24\textwidth, height=0.22\textwidth, trim={6pt 0 0 0}, clip]{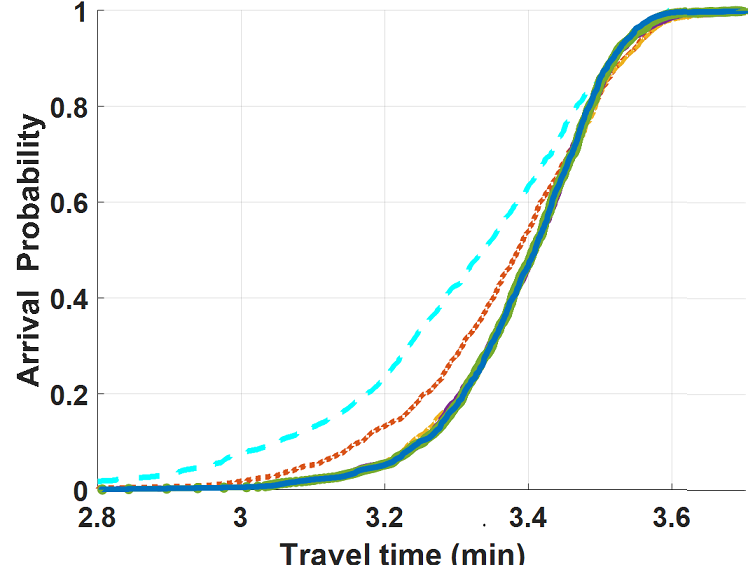}
\includegraphics[width=0.24\textwidth, height=0.22\textwidth, trim={6pt 0 0 0}, clip]{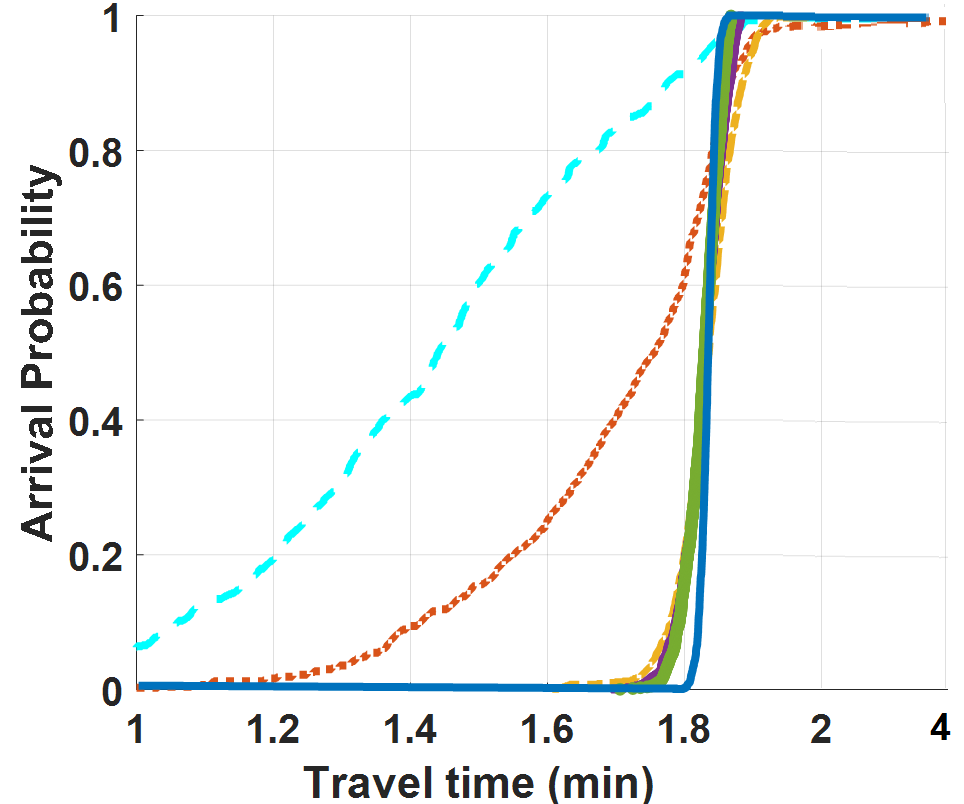}\\
\hspace{0.4in} Experiment 3 \hspace{0.8in}  Experiment 4\\
\includegraphics[width=0.24\textwidth, height=0.22\textwidth, trim={6pt 0 0 0}, clip]{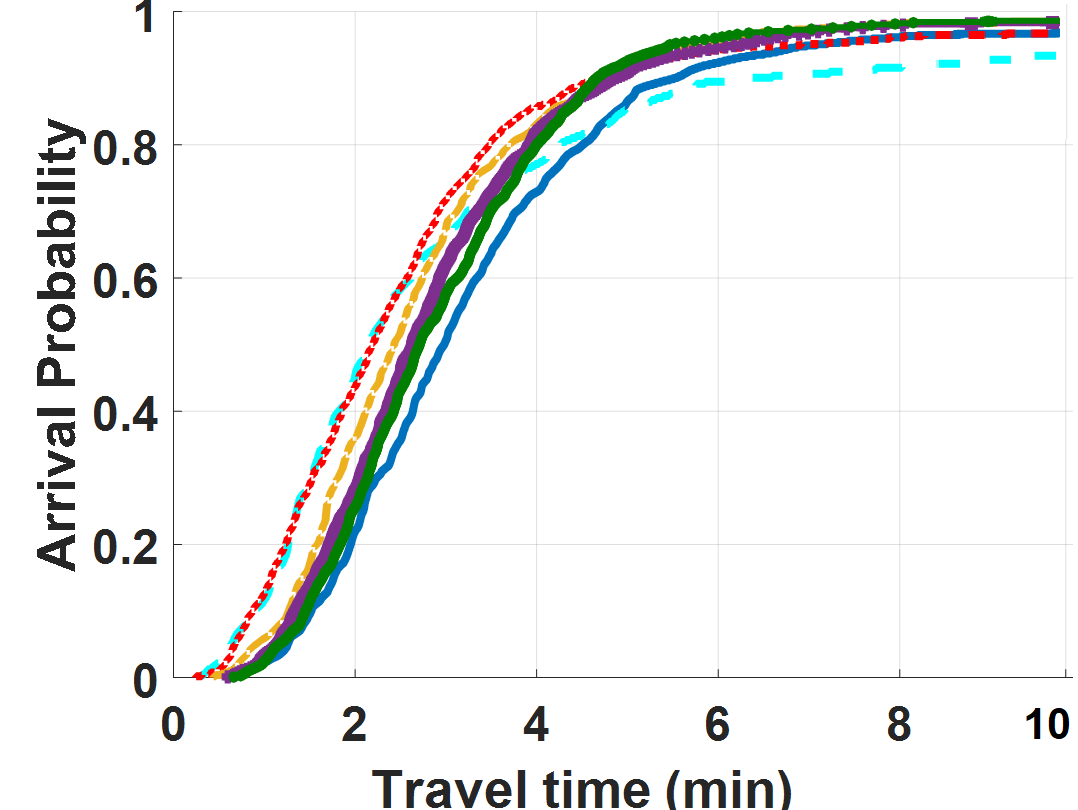}
\includegraphics[width=0.24\textwidth, height=0.22\textwidth, trim={6pt 0 0 0}, clip]{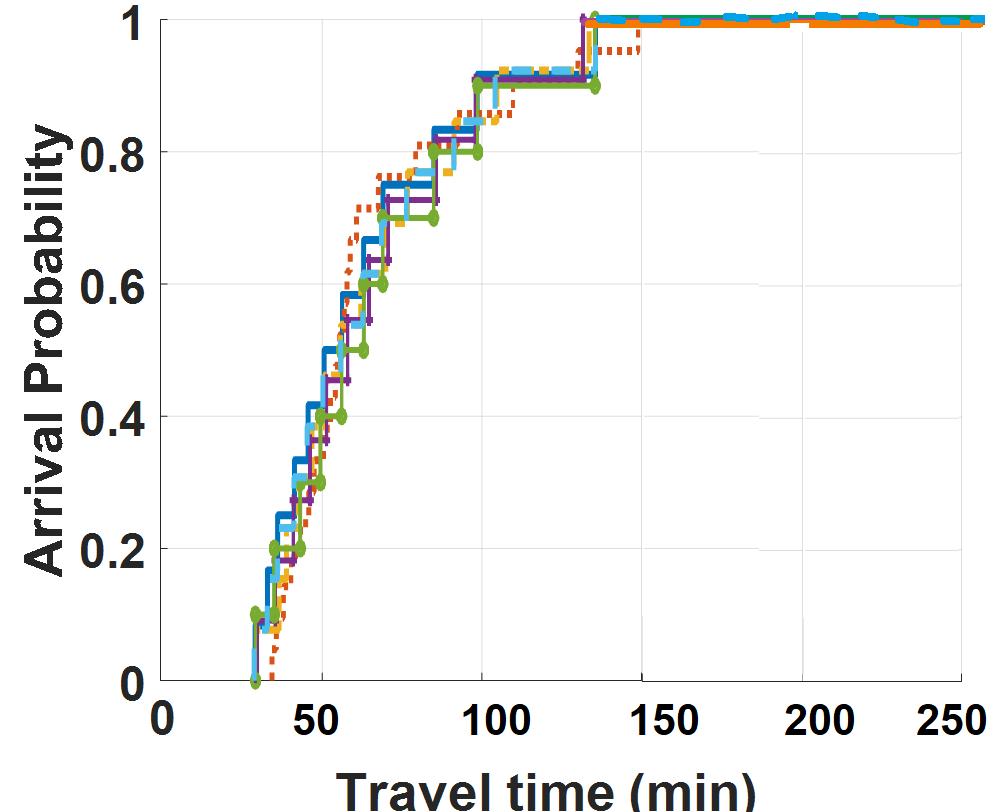}\\
\vspace{3pt}
\includegraphics[width=0.4\textwidth]{CDFlegend.png}\vspace{-6pt}
\caption{Weighted PP-TPU} \label{fig:weight}
\end{figure}

\vspace{-5pt}
Fig \ref{fig:CDF} presents the empirical CDFs of the privacy-preserving travel times in the four experiments.  As expected, the sanitization deviates the travel time distribution $\hat{f}^*(t)$ from the original $\hat{f}(t)$; the smaller per-trajectory privacy cost $\epsilon$ is, the more deviation there is. At $\epsilon\ge0.3$, $\hat{f}^*(t)$ is close to $\hat{f}(t)$ and satisfactory  utility can be reached for PP-TPU in all experiments.  From the CDF curves, we can read how quickly a traveler arrives at the destination with a certain level of confidence, and vice versa. For example, in Experiment 4, there is an 80\% probability that a traveler finishes the trip AB within 100 minutes if $\epsilon=0.5$ is used.  In addition to the unweighted TPU in Fig \ref{fig:CDF}, we also performed the weighted TPU analysis; the results are presented in Fig \ref{fig:weight}. A similar overall trend across $\epsilon$ is observed as in the non-weighted setting. In experiments 1 and 2, the weighting seems to affect $\hat{f}^*(t)$ more for smaller $\epsilon$, and the left tail of $\hat{f}^*(t)$ (smaller $t$) is more sensitive to the weighting than the right tail. In experiments 3 and 4, the weighted distributions are similar to the unweighted version across all $\epsilon$.

Table \ref{tab:Keff} presents the effective number of mapped full trajectories $K_{\text{eff}}$.   Due to the inherent error of the mapping algorithm, not every GPS record can be mapped onto the actual route where the traveler is on, or yield a sensible trajectory after mapping. Therefore, $K_{\text{eff}}$ is smaller than the number of trips even without any GI sanitization. With the GI sanitization and as $\epsilon$ decreases, $K_{\text{eff}}$ further decreases, as expected.
\begin{table}[!htb]\vspace{-6pt}
\caption{Effective number of mapped full trajectories $K_{\text{eff}}$}\label{tab:Keff}\vspace{-6pt}
\resizebox{1\columnwidth}{!}{\centering
\begin{tabular}
{c@{\hspace{0.5\tabcolsep}}|c@{\hspace{0.5\tabcolsep}}| c@{\hspace{0.5\tabcolsep}}c@{\hspace{0.9\tabcolsep}} c@{\hspace{0.9\tabcolsep}}c@{\hspace{0.9\tabcolsep}} c@{\hspace{0.9\tabcolsep}}|c@{\hspace{0.5\tabcolsep}}| c@{\hspace{0.9\tabcolsep}}c@{\hspace{0.5\tabcolsep}}}
\hline
&  &\multicolumn{5}{c|}{$\epsilon$} & original (no & \# trips\\
\cline{3-7}
& experiment & 0.05& 0.1 & 0.3 & 0.5 & 0.8 & sanitization) &\\ 
\hline 
&  1 & 792 & 853 & 873 & 889 & 892&901&1,000\\
&  2 & 682 & 721 &820  &834   & 845 & 876 &1,000\\
$K_{\text{eff}}$ &  3 &  229 & 314 &435  &460   & 478 & 513 &1,000\\
& 4 & 45 & 49 &52  &52   & 53 & 53 &419\\
\hline
\end{tabular}}\vspace{-6pt}
\end{table}

In summary, we can draw the following conclusions from the utility analysis in this subsection. (1) The quality of the PP-TPU analysis relates to the type and structure of the road network onto which the GPS records are mapped; some road networks are more sensitive to $\epsilon$ than others in the utility of sanitized trajectories. (2) The difference between the unweighted and weighted TPU analysis diminishes as  $\epsilon$ increases. (3) The CDFs of the privacy-preserving travel time in the 4 experiments are similar to  the original CDFs with the per-trajectory $\epsilon$ as small as $\approx0.3$, so is the effective number of mapped full trajectories, implying useful TPU analysis can be achieved with satisfactory privacy guarantees.

\vspace{-12pt}\subsection{Adversary Error}\vspace{-3pt}
Table \ref{tab:adversary} shows the expected AD between a sanitized and its original mapped trajectories calculated via Eq (\ref{eqn:AD}). Note that the 100 repeats were generated differently for experiments 1 and 2 vs. experiments 3 and 4 because the former two are synthetic data while the latter two  are quasi-real and real datasets, respectively. Specifically,  in experiments 1 and 2, we generated 100 GPS data sets per the simulation setting in Sec \ref{sec:setting}; in experiments 3 and 4, the 100 repeats were obtained by performing 100 sets of sanitization on a fixed GPS dataset in each experiment. As a result, the variability of AD  comes from two sources -- sampling error and sanitation error -- in experiments 1 and 2 and contains only the sanitization error in experiments 3 and 4.

The first observation is that the smaller $\epsilon$ is, the larger the distance is, as expected. Second, the AD value varies across the experiments for the same $\epsilon$, which makes sense, as the AD works with the distance between a pair of locations on a map and the road network matters. Given that the road networks differ in the four experiments, it is not surprising that the AD varies by experiment. Third, the adversary error measured by the AD at $\epsilon\le0.3$ is sufficiently large per location on a trajectory for each experiment ($\ge30$ meters). 
\begin{table}[!htb]
\caption{Mean (SD) average distance between mapped locations on sanitized and original trajectories (100 repeats)}\label{tab:adversary} \vspace{-6pt}
\resizebox{1\columnwidth}{!}{\centering
\begin{tabular}
{c@{\hspace{0.2\tabcolsep}}|c@{\hspace{0.2\tabcolsep}}| c@{\hspace{0.1\tabcolsep}}c@{\hspace{0.5\tabcolsep}} c@{\hspace{0.1\tabcolsep}}c@{\hspace{0.9\tabcolsep}} c@{\hspace{0.9\tabcolsep}}c@{\hspace{0.9\tabcolsep}} c@{\hspace{0.9\tabcolsep}}c@{\hspace{0.1\tabcolsep}}}
\hline
&  &\multicolumn{5}{c}{$\epsilon$}  \\
\cline{3-7}
& experiment & 0.05& 0.1 & 0.3 & 0.5 & 0.8 \\ 
\hline 
&  1 &180 (6.7) &87 (2.0) &30 (0.9)  & 18 (0.3) &11 (0.1) \\
AD$^\dagger$  &  2 &1189 (16.4) &814 (6.8) &435 (4.9)  & 341 (4.5) &296(2.4)\\
(meters)&  3 & 739 (24.1)& 438 (32.3) &185 (31.4)  & 121 (5.4) &99 (2.1)\\
& 4 & 214 (21.5) &108 (11.4)  &38 (3.7)   & 23 (2.1)  &13 (1.4)\\
\hline
\end{tabular}}
\end{table}

\begin{figure}[!htb]
\centering\vspace{-9pt}
\includegraphics[width=0.48\textwidth,height=0.5\textheight]{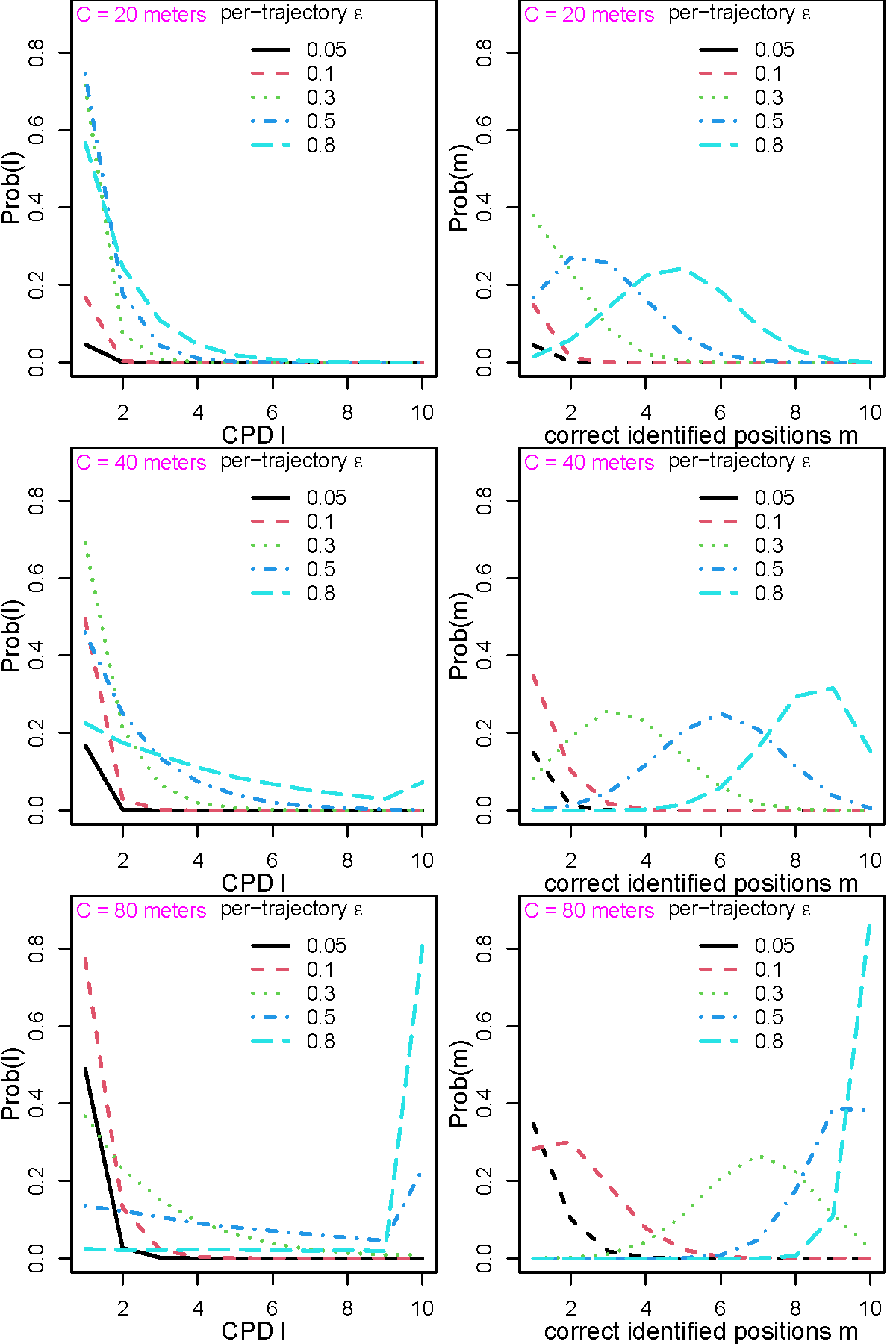}\vspace{-3pt}
\caption{Probability distributions of CPD $l$ (left column) and correctly identified positions $m$ (right column)} \label{fig:cpd}\vspace{-9pt}
\end{figure}

Fig. \ref{fig:cpd} presents the probability distributions of CPD $l$ and the correctly identified positions $m$ (whether consecutive or not) for three different clip radius $C$  (20, 40, and 80 meters) when the number of records per trajectory $n=10$ for different $\epsilon$. Since all 4 experiments used the same $n$ and $\epsilon$ value, the results in Fig. \ref{fig:cpd} apply to all four experiments. As expected, as $C$ increases (the criterion for claiming correct positioning loosens) or as per-trajectory $\epsilon$ increases, the adversary's accuracy for correctly identifying more positions and more consecutive positions increases. In the case of $C=80$ meters -- a rather relaxed criterion for correct identification, the probability of identifying 10 positions out of 10 is $>80\%$. The probability decreases to $\sim10\%$ for $C\!=\!40$ meters and  $\sim0\%$ for $C\!=\!20$ meters.  The plots also illustrate the differences between CPD $l$ and the number of correctly identified locations $m$. For example, for $C=20$, $\Pr(l\!=\!6)$ is close to 0\%, but $\Pr(m\!=\!6)$ is $\sim20\%$, regardless of whether the 6 positions are consecutive or not.

In summary,  we  can draw the following conclusions from the adversary error analysis in this subsection. (1) The magnitude of the adversary error closely relates to the road network type and structure. (2) The adversary error in reconstructing a trajectory from the sanitized trajectory around $\epsilon\le0.3$ is sufficiently large per the measures of AD and CPD to not pose serious privacy threats. (3) Taken together with the observations in the utility analysis,  a good trade-off between the PP-TPU utility and privacy protection can be achieved at per-trajectory $\epsilon\approx0.3$ with $\!\le\!10$ GPS records per trajectory in these 4 experiments. Users of the PP-TPU procedure can run similar analysis and choose an $\epsilon$ that leads to a good balance between utility and privacy protection for their specific problems.  

\vspace{-9pt}\section{Conclusions}\vspace{-6pt}
This paper addresses privacy-preserving TPU analyses. We employ the notation of GI to protect individual GPS spatial-temporal records and the subsequent TPU analysis. The proposed PP-TPU procedure can be adopted by service providers (e.g., mobile phone companies, GPS navigator apps) at the GPS data collection stage. We define the effective number of mapped full trajectories, the  usefulness concept, and different types of deviations in distance measures based on sanitized GPS records to quantify the utility of the sanitized trajectories. We also propose the concepts of average distance and consecutive positioning degree to assess the adversary error based on released GPS trajectory records.  Our analytical results and empirical studies suggest that it is feasible to employ the GI concept to collect and release GPS information for TPU analysis while guaranteeing location privacy for the individuals who contribute their GPS data.  Our future work will look into incorporating the dependency among the location points on the same travel trajectory and better utilizing the public road network maps to develop new randomization mechanisms of better utility without comprising privacy.

\vspace{-12pt}
\section*{Acknowledgments}\vspace{-3pt}
Fang Liu is supported by NSF Grant \#1717417 and Dong Wang is supported by the  China Scholarships Council program (NO. 201906270230) and NSFC Grant \#41971407. We also thank the editor, associate editor, and five reviewers for their useful comments and suggestions on the manuscript.

\vspace{-12pt}
\bibliographystyle{IEEEtranN}

\end{document}